\documentclass{aa}
\usepackage{graphicx}
\usepackage{natbib}
\usepackage{txfonts}
\def\etal{{\it et al.}}

\def\ie{{\it i.e.}}

\def\Rv{\lower 1.ex\hbox{$\buildrel R\over\sim$}}
\def\r2{$\rho^{2}$}
\def\r2s{\rho^{2}}
\def\r52{$\rho^{5/2}$}
\def\r52s{\rho^{5/2}}

\def\ea{e_{\alpha}}

\def\eb{e_{\beta}}
\def\ra{\rho_{\alpha}}
\def\rb{\rho_{\beta}}

\def\lova1{{l_{\alpha\alpha}\over a^{3}_{\alpha}} }
\def\lovb{{l_{\beta\beta}\over a^{3}_{\beta}} }
\def\lovc{{l_{\gamma\gamma}\over a^{3}_{\gamma}} }
\def\lovd{{l_{\delta\delta}\over a^{3}_{\delta}} }
\def\lova2{{l_{\alpha\alpha}\over a^{2}_{\alpha}} }
\def\lovb2{{l_{\beta\beta}\over a^{2}_{\beta}} }
\def\lovc2{{l_{\gamma\gamma}\over a^{2}_{\gamma}} }
\def\lovd2{{l_{\delta\delta}\over a^{2}_{\delta}} }

\def\r{{\vec r}}

\def\a{{\alpha}}

\def\b{{\beta}}

\def\apj{{\rm ApJ}}
\def\apjl{{\rm ApJL}}
\def\aanda{{\rm A\&A}}
\begin{document}
   \title{The Solar-Interior Equation of State with the Path-Integral Formalism}

   \subtitle{I. Domain of Validity}

   \author{Asher Perez\inst{1,2}, Katie Mussack\inst{3,4},
   Werner D\"appen\inst{3} \and Dan Mao\inst{3}}

   \offprints{Asher Perez}

   \institute{Laboratoire de Physique Th\'eorique (UMR CNRS/ULP 7085),
              Universit\'e Louis Pasteur de Strasbourg, 67084 Strasbourg Cedex,
              France
         \and Department of Applied Physics, Jerusalem College of
              Technology, 92221 Jerusalem, Israel
         \and Department of Physics and Astronomy, USC, Los Angeles,
              CA 90089-1342, U.S.A.
         \and Institute of Astronomy, University of Cambridge, Cambridge, CB3 0HA, UK\\
   \email{perez@lpt1.u-strasbg.fr,mussack@ast.cam.ac.uk,dappen@usc.edu,dmao@usc.edu}}

  \date{Received ; accepted }


\abstract
{}
{This is the first paper in a series that deals with solar-physics applications of
the equation-of-state
formalism based on the formulation of the so-called ``Feynman-Kac (FK)
representation''. Here, the
FK equation of state is presented and adapted for solar applications. Its
domain of validity is assessed. The practical
application to the Sun will be dealt with in Paper II. Paper III will extend the
current FK formalism to a higher order.}
{A recent rigorous quantum-statistical formalism for
Coulomb systems is used to compute the thermodynamical quantities for
solar modeling, taking into account the necessary requirements on smoothness and
accuracy. The FK formalism being a virial expansion, it suffers
from the well-known deficiency that it is limited to nearly full ionization.
This point is elaborated in detail, and the quantitative criterion for
the domain of validity of the FK equation of state is established.}
{Use of the FK equation of state is limited to physical conditions for which
more than 90\% of helium is ionized. This incudes the inner
  region of the Sun out to about .98 of the solar radius. Despite this limitation, in the
parts of the Sun
where it is applicable, the FK equation of state has the power to be
more accurate than the equations of state currently used in solar modeling.
The FK approach is especially suited to study physical effects such as
Coulomb screening, bound states, the onset of recombination of fully ionized
species,
as well as diffraction and exchange effects.}
{Despite technical difficulties in its application, there are unique features in the FK
approach that promise to turn it into the most exact of the available formalisms, provided
FK is restricted to the deeper layers of the Sun where more than 90\% of helium is
ionized. The localizing power of helioseismology allows a test of the
FK equation of state. Such a test will be beneficial both for better
solar models and for tighter solar constraints of the equation of
state.}


\authorrunning{Perez et al.}
\titlerunning{Equation of State with the Path Integral Formalism I.}

\maketitle
%

\section{Introduction}

This paper is the first of a series of three devoted to the study of the
equation of state based on the formalism of the path integral in the
framework of the Feynman-Kac (hereinafter FK)
representation. This formalism leads to
a virial expansion of the thermodynamic functions in powers of the
total density of a Coulomb plasma \citep{ap92,acp94,acp95,ap96}. The
path-integral approach is the latest in a series of attempts to improve
the equation of state for the solar interiors. However, so far there
has only been a very limited amount of astrophysical applications
of that formalism. One \citep{pc04}
was an application to the screening enhancement of
nuclear fusion. Another one
was a precursor \citep{pd95} of the present series of
articles, which addressed the possibility of using the
FK formalism in solar and stellar modeling. Here, we embark on a more
systematic application of the FK
representation. It is well known that solar and stellar modeling requires
thermodynamic quantities that are smooth, consistent, valid over a
large range of temperatures and densities, and incorporate the most
important astrophysically relevant chemical elements. Let us briefly
review the two major
equation-of-state efforts of the last 20 years. Both were made in the
context of the two most recent opacity recalculations.

One of them is the international Opacity Project
\citep[OP; see the books
by][]{sea95,ber97}; it contains the
so-called Mihalas-Hummer-D\"appen equation of state
\citep[][hereinafter MHD]{hm88,mdh88,dmh88,ndhm99,tr06}, dealing
with {\it heuristic} concepts about the modification of atoms and
ions in a plasma. Specifically, in MHD the destruction of atomic
states is modeled, on the one hand by neutral species according to
excluded volume, and on the other hand by Stark dissolution due to
the surrounding microfield. From the included 15 astrophysically relevant
elements (H, He, C, N, O, Ne, Na, Mg, Al, Si, S, Ar, K, Ca, Fe),
users can select any subset. Although MHD in its current version
does not contain free parameters, several of the physical
quantities are, in principle, open to revision. Possible candidates for
adjustment are (i) the atomic and ionic radii (which can be chosen to match
observations), (ii) the strength of the assumed microfield
distribution, and (iii) the specific form of the interaction potential
between species. Since MHD is basically a heuristic equation of state,
such future improvements will be legitimate and they can be developed
under the guidance of observations.
For more details see~\cite{tr06}.

The other recent opacity effort is the OPAL project pursued at
Livermore.
In contrast to OP, its underlying equation of state has relied on
the so-called {\it physical picture}, which is built on the notion
of fundamental particles only, that is, electrons and nuclei. It
provides a systematic method to include nonideal effects. Its
so-called ACTEX (``activity expansion'') equation of state
\citep[see][and references therein]{ro86,rsi96,rn02} became part of
the OPAL opacity project at Livermore
\citep{irw87,ir91,ir93,ir95,ir96}. For this reason, the ACTEX
equation of state is hereinafter referred to as the OPAL equation of
state. As mentioned, OPAL is a physical-picture formalism, starting
out from the grand canonical ensemble of a system of the basic
constituents (electrons and nuclei) interacting through the Coulomb
potential. In the currently released version, the OPAL equation of state
includes the following astrophysically relevant elements:
H, He, C, N, O, Ne (but the OPAL {\it opacity} calculation
includes other elements,
such as Fe, as well). In OPAL,
any effects of the plasma environment on the internal
states are obtained directly from the statistical-mechanical
analysis, rather than by assertion as in the chemical picture.
Therefore, in contrast to the intuitive MHD,
in OPAL there are no adjustable parameters.
On the one hand, this is its strength,
on the other hand, it means that one cannot easily upgrade OPAL
even if observations were to suggest so. Thus in the current
OPAL equation of state tables,
any residual inaccuracy is either
structural, a
result of (i) the finite truncation in the underlying
activity expansion, and (ii) the utilization of effective atomic and ionic
potentials, or it could be due to the limited number of
chemical elements in the current tables.

The equation of state for solar and stellar structure and seismology
has to be {\it formally} precise and consistent, even before the
question of the accuracy of the physical description is asked. It
has to satisfy four conditions: {\it i}) a large domain of
applicability (in $\rho$, ~$T$), {\it ii}) a high precision of its
numerical realization, {\it iii}) consistency between the
thermodynamic quantities, and {\it iv}) the possibility to take into
account relatively complex mixtures with at least several of the
more abundant chemical elements. These requirement are generally
easy to satisfy with relatively simple formalisms, such as the
popular, widely-used CEFF equation of state \citep{cd92}, which is an
offspring of the widely used Eggleton-Faulkner-Flannery (EFF)
equation of state \citep{eff73}, or the other offspring of EFF, the
SIREFF equation of state~\citep{rsi96}.

However, the demands from helioseismology are such that formal
considerations are not sufficient. The equation of state has
to be accurate, not just smooth. Even state-of-the-art
equations of state such as OPAL and MHD are not completely sufficient
\citep{da06}, and improvements are still necessary. Improvements can
be made either formally (phenomenologically) or rigorously. A formal option is
to merge the good features of OPAL and MHD
\citep{sta03,liad03,lia04,da09} to create an equation
of state that matches reality as closely as possible.
However, as a phenomenological effort, such a hybrid formalism would
not be a scientific improvement but merely a better
computational tool for solar and stellar
modeling. As far as the development of a real theory is concerned,
genuine improvements in the physical model must be made.
The rigorous equation of state in the FK
representation is such an improvement.
In principle, it is a better theory than either MHD or OPAL.
A case in point is the treatment of Coulomb screening. Here, the electrons are
taken into account with the exact quantum-mechanical
exchange and diffraction terms. These are corrections that deal, on the one hand,
with the ``Pauli blocking'' of electrons (exchange terms) and, on the other hand,
their finite thermal de Broglie wavelength (diffraction terms). For these terms,
both MHD and OPAL use approximations.
However, this better physics of FK comes with a price: the FK
equation of state has a limited domain of
validity unlike the more global MHD and OPAL. It only works for nearly fully ionized plasmas.

As a theory, FK is quite similar to OPAL, since it
is realized in the physical picture and the grand canonical ensemble. However,
its net result is that of
a virial (density) expansion (unlike OPAL, which is based on activity).
The virial coefficients of the FK expansion
are evaluated by the
path-integral formalism, which in its essence uses
the equivalence between a point-charge quantum system and a
classical one made of extended filaments. In this spirit, the
calculation of all non-ideal contributions to the equation of state becomes
possible,
systematically, exactly, and analytically.
The fact that FK has the form of a
virial expansion is the source of its limitation to
near full ionization. In Section~2 we illustrate the well-known fact that in the case
of reacting gases, virial expansions are inadequate to deal with recombination (unlike
the activity expansions of OPAL). We discuss this deficiency quite in detail and develop
a quantitative criterion for the
validity of the FK equation of state. The adopted criterion is a limitation to physical conditions
such that more than 90\% of helium is ionized. The subsequent sections discuss
the physical condition quantitatively, in terms of a solar model.

Despite the aforementioned limitation to nearly fully-ionized
matter, the FK approach can be applied to solar modeling. The
reason for this is twofold. First, a large part of the solar
interior (out to about .98 of the solar radius) lies inside its domain of
validity, and second, helioseismology
allows a {\it localized} analysis of
thermodynamic properties despite the fact that there is
no complete coverage of the Sun by the FK equation of state.

Here the FK equation of state is presented and its domain of validity
discussed. Paper II will then be dedicated to solar
applications. They will be based on ``mixed'' thermodynamic tables, with
FK in the highly-ionized part, and a conventional equation of state (OPAL or MHD)
outside the domain of validity of FK. We note in passing that once these tables
exist they can be used also for stellar modeling, but it is not clear if
stellar application would ever require such an accuracy of the equation of state, in
contrast to solar application for which this need has been well established~\citep[see, {\it e.g.},][]{da06}. In any case, stellar applications are beyond
the scope of this current series of papers.
Finally, paper III will be a natural extension of the current level of the FK
formalism. While the current level of FK is based on an expansion of
pressure in terms of the density with up to
order $\rho^{5/2}$, the next (rather elaborate) step is the
extension to calculation of the thermodynamic functions
up to order $\rho^3$. Then we will be able to take into account
three-body effects, allowing us to treat helium exactly,
beyond the hydrogenic approximation.




\section{Illustration of the limitation of virial expansions for reacting systems}

The principle can be understood by considering the exact equation of the
ideal (non-relativistic) Fermi gas. One starts out from
an {\it implicit} equation of state,
obtained from the grand-canonical
partition function \cite[see, e.g.][]{hu63}

\begin{equation}
{p \over {kT}} = {1 \over {\lambda^3}} f_{5/2}(\tilde z), \ \ \
{N \over V} = {1 \over {\lambda^3}} f_{3/2}(\tilde z)\ .
\label{E.01}
\end{equation}
Here, $p,T,N,V,k,\lambda$ are pressure, temperature, particle number, volume,
Boltzmann constant,
and the thermal de Broglie length $\lambda = h / \sqrt{2\pi mkT}$, respectively.
The fugacity $\tilde z = \exp(\mu / (kT))$ ($\mu$ being
the chemical potential) is defined implicitly by Eq.~(1), by virtue of
the Fermi integrals

\begin{equation}
f_{5/2}(\tilde z) = {4 \over {\sqrt{\pi}}}\int_0^\infty dx\ x^2\ln(1 + \tilde ze^{-x^2}),\
\label{E.02}
\end{equation}

\begin{equation}
f_{3/2}(\tilde z) = \tilde z{d \over {d\tilde z}}f_{5/2}(\tilde z)\ .
\label{E.03}
\end{equation}
Expanding the Fermi integrals as a series in the high-temperature limit, and
eliminating the fugacity $\tilde z$, one obtains the virial (density) expansion

\begin{equation}
{pV \over {NkT}} = 1 + {\lambda^3 \over {2^{5/2}}}{N \over {V}} + O({N \over V}^2) + ...\ ,
\label{E.04}
\end{equation}
Alternatively, one can continue to work with fugacity expansion

\begin{equation}
{p \over {kT}} = {1 \over {\lambda^3}}\left[{\tilde z} - {1 \over
{2^{5/2}}}{\tilde z}^2 + ... \right],
\label{E.05}
\end{equation}

\begin{equation}
{N \over V} ={1 \over {\lambda^3}}\left[{\tilde z} - {1 \over
{2^{3/2}}}{\tilde z}^2 + ... \right].
\label{E.06}
\end{equation}
With the introduction of the {\it activity} $z = {\tilde z} / {\lambda}^3$
we obtain an even more convenient expansion

\begin{equation}
{p \over {kT}} = {z} - {\lambda^3 \over
{2^{5/2}}}{z}^2 + ...\ ,
\label{E.07}
\end{equation}

\begin{equation}
{N \over V} = {z} - {\lambda^3 \over
{2^{3/2}}}{z}^2 + ...\ ,
\label{E.08}
\end{equation}
These expressions above are exact for the partially degenerate
ideal Fermi gas, they retain their formal structure for any real gas (this
follows from their origin in the grand-canonical partition function). Therefore
all the physics lies in the value of the coefficients of the expansions. For
the activity, one formally writes

\begin{equation}
{p \over {kT}} = {z} - b_2{z}^2 + ...\ ,
\label{E.09}
\end{equation}

\begin{equation}
{N \over V} = {z} - 2b_2{z}^2 + ...\ ,
\label{E.10}
\end{equation}
and for density,

\begin{equation}
{pV \over {NkT}} = 1 + B_2 {N \over {V}} + O({N \over V}^2) + ...\ .
\label{E.11}
\end{equation}
Inspection of Eqs.~(\ref{E.04}) and~(\ref{E.07}) reveals the relation

\begin{equation}
B_2 = - b_2,
\label{E.12}
\end{equation}
While we derived the relation for the ideal Fermi gas, Eq.~\ref{E.12}
is a general result, valid for all real
gases~\citep[see, {\it e.g.}][]{hu63}.
The second virial coefficient is a function of temperature only; an explicit
expression is obtained from
classical statistical mechanics by the
two-body configuration integral over the interaction potential $U_{12}$

\begin{equation}
B_2 = {1 \over 2}\int (1 - e^{-U_{12}/(kT)})\ dV
\label{E.13}
\end{equation}

\begin{figure}
   \includegraphics[width=0.45\textwidth]{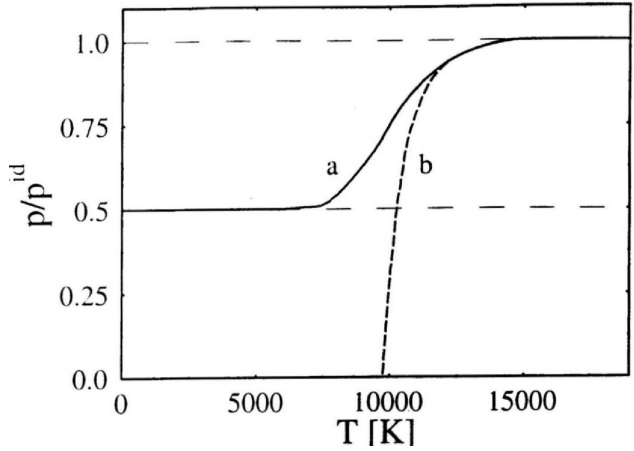}
   \caption{Illustration of the behavior of activity (solid line) and density expansions
   (dashed line) for reacting systems, here for the simple reaction of neutral hydrogen H
   recombining to hydrogen molecules ${\rm H}_2$. $P_{\rm id}$ refers to
   fully dissociated hydrogen. The activity expansion (a) correctly describes
   recombination with just one non-ideal term (equations \ref{E.09},\ref{E.10}).
   Without further high-order terms, the corresponding virial expansion (b)
   (\ie the reacting analog to the single-species equation \ref{E.11}) exhibits
   catastrophic
   non-physical behavior
   (negative pressure!)~\citep{ebe76}. The case of ionization is analogous.
   \label{fig:viract}}
 \end{figure}

Of course, as {\it infinite} series, activity and density expansions would be
strictly equivalent, but when dealing with recombination,
a density expansion would have to go to a very
high order to match the accuracy of even a second-order activity expansion.
Indeed, for recombinations, the activity expansion~(\ref{E.09},\ref{E.10})
is vastly superior to the virial expansion~(\ref{E.11}). This is easily seen from
Eq.~(\ref{E.13}), where for a (necessarily) attractive two-body potential and
low temperatures $B_2$ becomes large negative, which will cause inevitably a
negative total pressure at some low temperature.
In other words, the virial expansion breaks down.
This breakdown is illustrated by the dashed line of
Figure~\ref{fig:viract} \citep[from][]{ebe76}. For simplicity,
that figure is for hydrogen molecular
dissociation, but qualitatively it would be the same for ionization.
In contrast, the activity expansion can avoid this breakdown, as revealed by
a more detailed discussion of Eqs.~(\ref{E.09},\ref{E.10}) that can be
found elsewhere~\citep[~{\it e.g.}][]{ebe76,kre05}. Importantly,
already to lowest non-ideal
order, the activity expansion can correctly describe
complete low-temperature recombination (solid line in
Figure~\ref{fig:viract}). This property of activity expansions is the principal reason for
the remarkable success of the
OPAL equation of state.

In contrast, as a virial expansion,
the FK equation of state suffers from the pathology
illustrated by Figure~\ref{fig:viract}, irrespective of the fact that its low-order
virial coefficients are exact. Nevertheless there is a domain of applicability
where we can use FK: Figure~\ref{fig:viract} shows the overlap region of the
virial and activity expansions. A
natural quantitative limit
for the application of a virial expansion becomes apparent: as long
as the physical conditions are such that more than about
90\% of the bound particles remain ionized, the resulting
thermodynamic quantities still will be reliable.
In the following we adopt this criterion for the FK approach.

\section{The Path-integral formalism in the framework of the
Feynman-Kac representation.}

In many situations, stellar matter can be adequately
described in terms of
quantum plasmas made of electrons and nuclei. In this part, we
summarize the main steps of the path integral formalism in the FK
representation.
Details of the calculation can be found in \citep{ap92,acp94,acp95}.
The general method can be
divided in two steps. In a first step only Maxwell-Boltzmann
(MB) statistic is considered and all exchange effects are omitted.
Using the FK
representation, an equivalent classical system
made of closed filaments interacting via {\it two-body}
forces is introduced. Applying traditional techniques of classical mechanics
inspired by Abe-Meeron's method \citep{ab1959,me1958},
a formal diagrammatic representation of the MB quantities was obtained, which is
term-by-term well-behaved.

In a second step of this approach, the exchange contribution
was introduced, evaluated
perturbatively via Slater sums. Use of the FK representation
leads to the appearance of
open filaments (impurities) immersed in a bath of closed
filaments described by MB statistics. This problem can be dealt with
using the method that is explained in the MB case to inhomogeneous
situations.

Collecting all the contributions arising from both the MB and the
exchange treatment, a systematic virial
expansion of the equation of state was obtained, which includes
screening, diffraction,
bound states and exchange contributions to the ideal gas (depending on the statistics
of the particles). The expression of
the pressure was calculated up to and including order $\rho^{5/2}$
(with $\rho$ being the overall density of the system). To this order of density,
the calculation of \cite{ap92,acp94,acp95} describes exactly
all the effects pertaining to
bound states, diffusive and exchange processes of a system consisting of
two species of elementary particles.

\subsection{Exact expression of the equation of state to the order of $\rho ^{5/2}$}

The virial expansion of the pressure to
order $\rho ^{5/2}$ gives \citep{ap92}

\begin{eqnarray}
\lefteqn{\beta P\ =\ \sum_{\alpha} \rho_{\alpha}-{ \kappa_{D}^{3}
\over 24 \pi}}\nonumber\\
&&+{\pi \over
6}(\ln{2}-1)\sum_{\alpha,\beta}\beta^{3}\ea^{3}\eb^{3}\ra\rb
\nonumber \\
&&-{ {\pi} \over {\sqrt{2}}} \sum_{\alpha,\beta} \rho_{\alpha}
\rho_{\beta}
  {\lambda^{3}_{\alpha\beta}} Q(x_{\alpha \beta})
-{\pi \over 3}\beta^{3}\sum_{\alpha,\beta} \ra\rb e_{\alpha}^{3}
e_{\beta}^{3}\ln{(\kappa_{D} \lambda_{\alpha \beta})}\nonumber\\
&&+{{\pi} \over {\sqrt{2}}}\sum_{\alpha} {(-1)^{2\sigma_{\alpha}+1}
\over
 (2\sigma_{\alpha}+ 1)} {\lambda^{3}}_{\alpha \alpha}\ra^{2}
E(x_{\alpha\alpha})\nonumber\\
&&-{3\pi \over{2\sqrt{2}}}\beta\sum_{\alpha,\beta}\ea\eb\kappa_{D}\ra\rb
{\lambda_{\alpha\beta}}^3 Q(x_{\alpha\beta})\nonumber\\
&&- {\pi \over 2}\beta^{4}
 \sum_{\alpha,\beta}\ra\rb
 e_{\alpha}^{4}e_{\beta}^{4}\kappa_{D}\ln{(\kappa_{D} \lambda_{\alpha \beta})} \nonumber\\
&&+{3\pi \over{2\sqrt{2}}}\beta\sum_{\alpha}
{(-1)^{2\sigma_{\alpha}+1} \over (2\sigma_{\alpha}+ 1)}
\lambda^{3}_{\alpha\alpha}\ra^{2}{e_{\alpha}^2}
\kappa_{D} E(x_{\alpha\alpha}) \nonumber\\
&&+{1 \over 16 }\sum_{\alpha}{{\beta^{2}\hbar^{2}\ea^{2}} \over
{m_{\alpha}}}
  \kappa_{D}^{3}\ra+{\pi}({1\over 3}-{3 \over 4}\ln{2}+{1 \over
  2}\ln{3})\times\nonumber\\
&&\sum_{\alpha,\beta}\beta^{4}\ea^{4}\eb^{4}\kappa_{D}\ra\rb  \nonumber \\
&&+C_{1}\sum_{\alpha,\beta,\gamma}\beta^{5}\ea^{3}\eb^{4}e^{3}_{\gamma}\kappa_{D}^{-1}\ra\rb\rho_{\gamma}\nonumber\\
&&+C_{2}\sum_{\alpha,\beta,\gamma,\delta}\beta^{6}\ea^{3}
\eb^{3}e^{3}_{\gamma}e^{3}_{\delta}\kappa_{D}^{-3}
\ra\rb\rho_{\gamma}\rho_{\delta} \label{E.14}
\end{eqnarray}

\noindent with
$\kappa _D = \displaystyle{{(4 \pi \beta \sum
_{\alpha} e^2 _{\alpha} \rho _{\alpha})}^{1/2}}$, $\lambda
  _{\alpha\beta} = {(\beta \hbar ^2 / m_{\alpha \beta})}^{1/2}$, $m_{\alpha
    \beta} = m_\alpha m_\beta / (m_\alpha + m_\beta)$, ${x_{\a\b}}=-\sqrt{2}
{l_{\a\b}}/ {\lambda_{\a\b}}$, $l_{\alpha \beta}
= \beta e_\alpha e_\beta$,  $C_1 = 15.205 \, \pm
\, .001$, $C_2 = - 14.733 \, \pm \, .001$, and the Euler-Mascheroni
constant $C =0.577216$.

In Equation~(\ref{E.14}), $Q ( x_{\a\b})$ refers to  the so-called quantum
second-virial coefficient first introduced by Ebeling and co-workers
\citep{ebe76,krae86}.
\begin{eqnarray}
\lefteqn{Q(x_{\a\b})=}\nonumber\\
&&{{1} \over {(\sqrt {2}{\pi \lambda^{3}_{\alpha \beta}})}}
 \lim_{R\rightarrow\infty}
\Biggl\{ \int_{r<R} d{\vec
r}\Bigl[(2\pi\lambda^{2}_{\alpha\beta})^{3/2}<{\vec r} | e^{-\beta h_{\alpha\beta}}|{\vec r}>-1 +\nonumber\\
&&{\beta e_{\alpha}e_{\beta} \over
r}-{\beta^{2}e_{\alpha}^{2}e_{\beta}^{2}
 \over 2r^{2}}\Bigr]+{2\pi \over 3}\beta^{3}e_{\alpha}^{3}e_{\beta}^{3}
[\ln ({{3 \sqrt{2}R} \over {\lambda_{\alpha \beta}}})+C]\Biggr\}
\label{E.15}
\end{eqnarray}

\noindent while $E (x_{\a\a})$ is defined as the two-body exchange integral,

\begin{equation}
E(x_{\a\a})=
(2{\sqrt {\pi}}) \lim_{R\rightarrow\infty} \int_{r<R} d{\vec
r}<-{\vec r}
 | e^{-\beta h_{\alpha\alpha}}|{\vec r}>
\label{E.16}
\end{equation}

In Equations~(\ref{E.15}) and~(\ref{E.16}), $h_{\alpha \beta}$
is the one-body Hamiltonian of the relative particle with mass
$m_{\alpha \beta}$ submitted to the Coulomb potential $e_\alpha
e_\beta /r$. The functions $Q$ and $E$ only depend on the temperature
via the single dimensionless parameter $(-l/\lambda)$.

\subsection{Deriving Other Thermodynamic Variables}
The expression for the pressure in Equation~(\ref{E.14}) can be used
  to derive other thermodynamic variables.
  For example, the relationship between the pressure and the free energy is given by
\begin{equation}
\beta P=
\sum_{\alpha} \rho_{\alpha} {{\partial (\beta f) }\over
{\partial \rho_{\alpha}}} - \beta f
 \label{E.17}
\end{equation}
where $f$ is the free energy per unit volume.
The internal energy per unit volume $u$ is then found as a
  derivative of the free energy
\begin{equation}
 {u} = {{\partial}\over{\partial\beta}}\left(\beta f \right)_\rho
\end{equation}
 Other
  quantities of interest can be derived in a similar way via the usual thermodynamic
  derivatives.

\subsection{The general structure of $\beta P$}

   \begin{figure}
   \centering
   \includegraphics[width=0.45\textwidth]{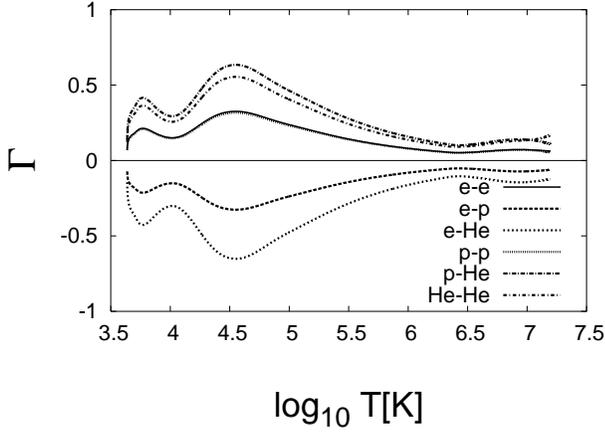}
   \caption{Coupling parameters for electrons,
   protons, and helium nuclei from the surface to the center of the
   Sun. \label{fig1}}%
   \end{figure}

   \begin{figure}
   \centering
   \includegraphics[width=0.45\textwidth]{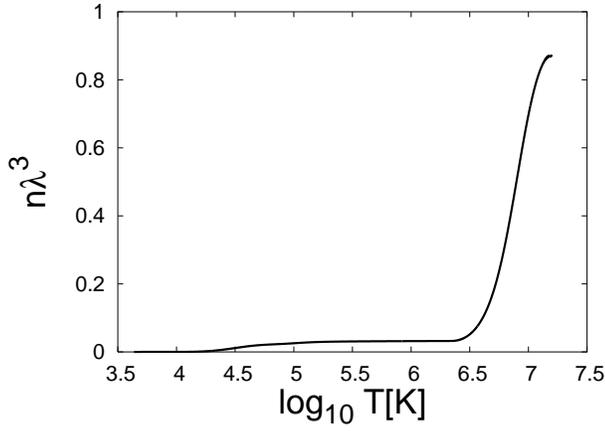}
   \caption{Degeneracy parameter for electrons from
    the surface to the center of the Sun. \label{fig2}}%
    \end{figure}

   \begin{figure}
   \centering
   \includegraphics[width=0.45\textwidth]{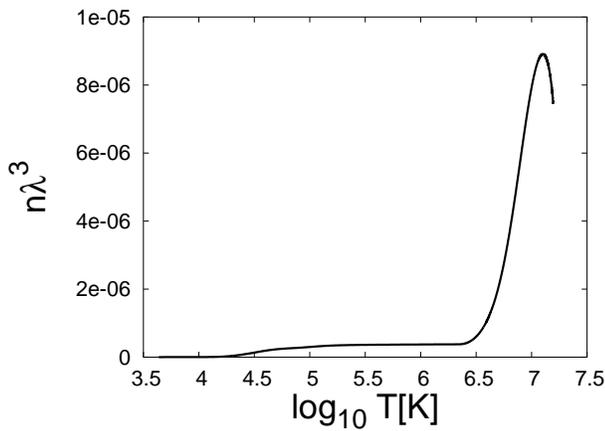}
   \caption{Degeneracy parameter for protons from
    the surface to the center of the Sun. \label{fig3}}%
    \end{figure}

   \begin{figure}
   \centering
   \includegraphics[width=0.45\textwidth]{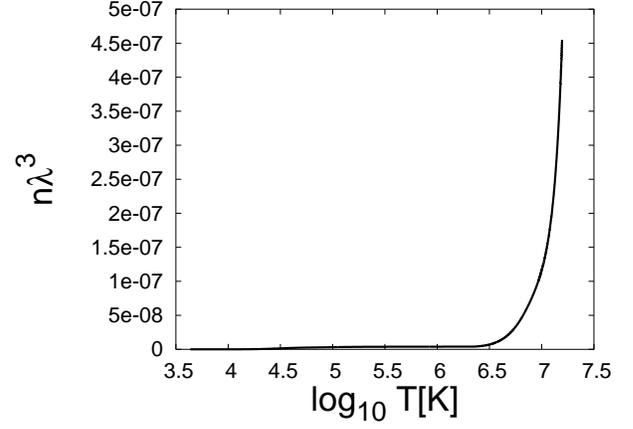}
   \caption{Degeneracy parameter for helium nuclei
    from the surface to the center of the Sun. \label{fig4}}%
    \end{figure}

   \begin{figure}
   \centering
   \includegraphics[width=0.45\textwidth]{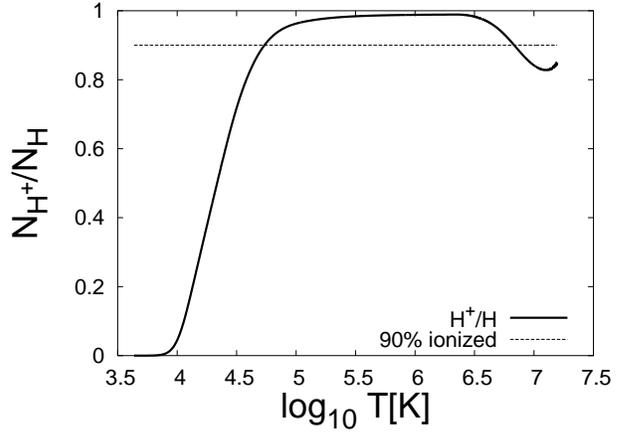}
   \caption{Ionization degree of hydrogen from the
    surface to the center of the Sun. \label{fig5}}%
    \end{figure}

   \begin{figure}
   \centering
   \includegraphics[width=0.45\textwidth]{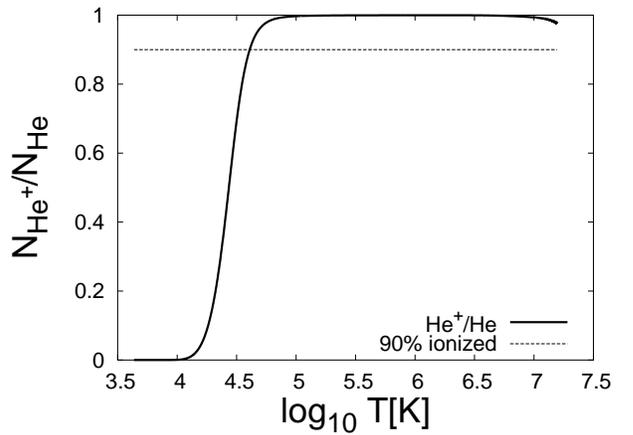}
   \caption{First ionization degree of helium from
    the surface to the center of the Sun. \label{fig6}}%
    \end{figure}

   \begin{figure}
   \centering
   \includegraphics[width=0.45\textwidth]{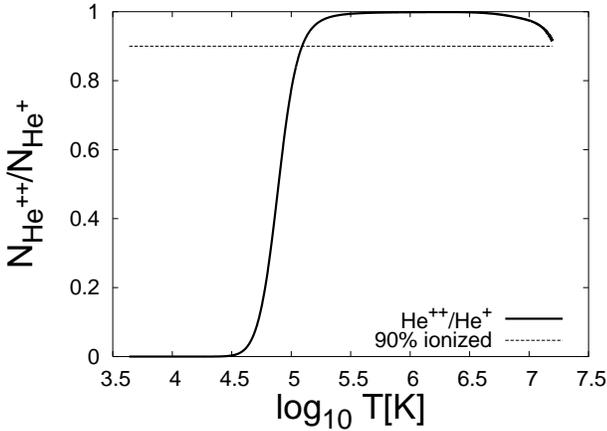}
   \caption{Second ionization degree of helium from
    the surface to the center of the Sun. \label{fig7}}%
    \end{figure}

   \begin{figure}
   \centering
   \includegraphics[width=0.45\textwidth]{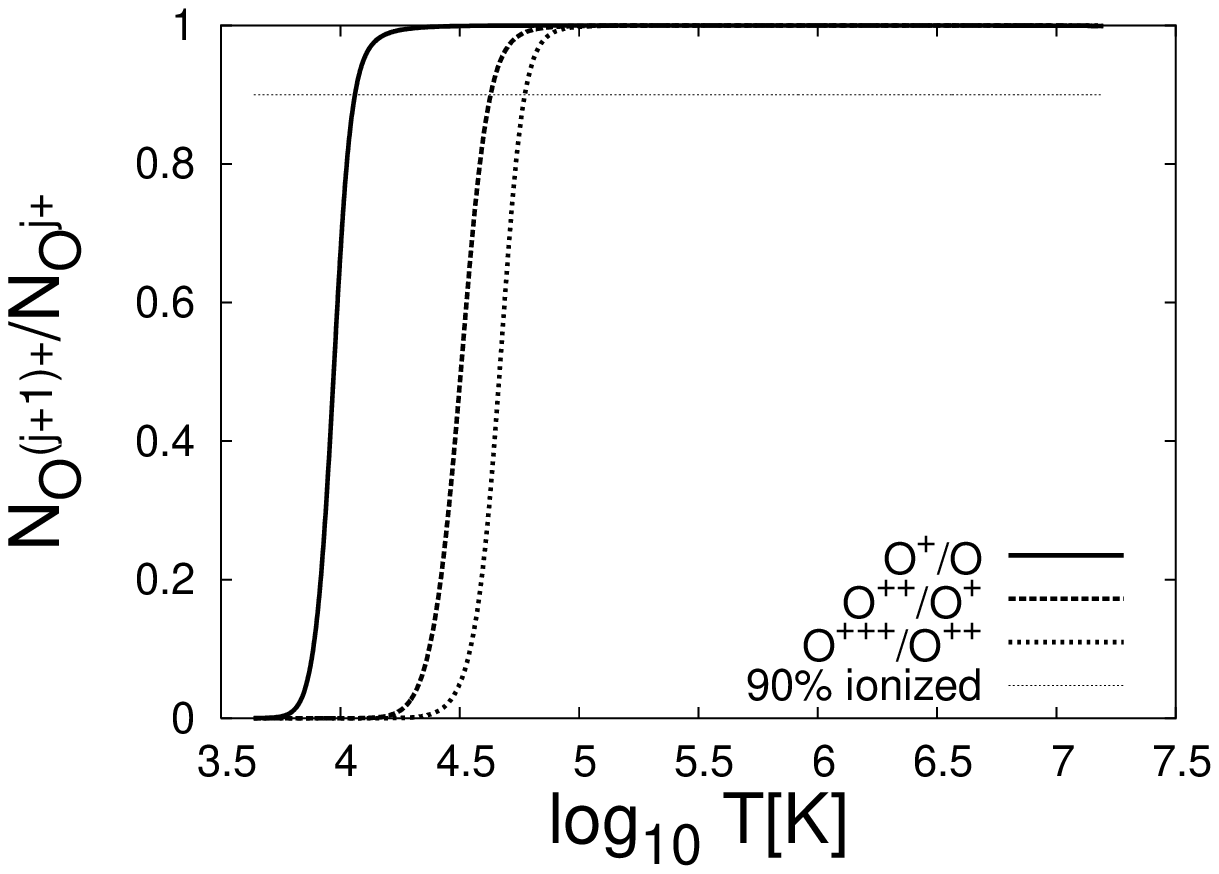}
   \caption{First and second ionization degree of
    oxygen (as a representative heavy element) from the surface to the
    center of the Sun. \label{fig8}}%
    \end{figure}

First, the classical contributions of the long-range part of the
interactions are polynomials in the inverse temperature, the charges
and the densities, which do not involve Planck's constant. The
involved coefficients are evaluated analytically or, like $C_1$ and
$C_2$, by numerical computations of dimensionless integrals. The
lower-order contribution is nothing but the familiar Debye-H$\ddot
u$ckel term in $\rho ^{3/2}$, and constitutes the leading correction
to the MB ideal pressure in Equation~(\ref{E.17}). It can be
quoted, here, that the classical OCP limit up to order $\rho^{5/2}$
included does coincide with those calculated by \citet{comu69}, as
it should.

The contribution of quantum diffraction at large distances appears
only at the order $\rho ^{5/2}$ and reduces to ${1 \over 16
}\sum_{\alpha}{{\beta^{2}\hbar^{2}\ea^{2}} \over {m_{\alpha}}}
\kappa_{D}^{3}\ra$ The occurrence of this term shows that the long-range
part of the interactions cannot be entirely treated at a classical
level. The diffraction correction is merely proportional to $\hbar$,
because it arises from large distances where the quantum effects can
be treated perturbatively ``$\grave a$ la-Wigner-Kirkwood".

The term $\rho _\alpha {\,} \rho _\beta {\,}\lambda ^3 _{\alpha
\beta}{\,} Q (-\sqrt{2}{{l_{\alpha \alpha}} \over {\lambda_ {\alpha
\alpha}}})$ is the total contribution from both bound and scattering
states of two charges $e_\alpha$ and $e_\beta$. The truncation of $<
\vec r{\,} \vert \exp \ ( - \beta h_{\alpha \beta}) {\,}\vert
{\,}\vec r >$ in the integral defining Q ensures that this
contribution is finite. This regularization is not an arbitrary
mathematical artifact. It is directly related to the truncated
structure of the bond $f_T$, and reflects the screening of the
Coulomb interaction at large distances. For opposite charges such
that $e_\alpha e_\beta < 0$, one may extract from Q a contribution
of the bound states which reduces to the familiar Planck-Larkin (PL)
sum

\begin{equation}
\sum_{n=1}^{\infty}n^2\Bigl[\exp(-\b
\epsilon_n^{\a\b})-1+\b\epsilon_n^{\a\b}\Bigr] \label{eqn:27}
\end{equation}

\noindent where ${\epsilon}^{\alpha \beta}_n = -e^2_\alpha
{\,}e^2_\beta {\,}m_{\alpha \beta}/ (2{\,}\hbar ^2 n^2)$ are the
energy levels of the hydrogenic atom with Hamiltonian $h_{\alpha
\beta}$. However, other definitions of the bound states
contributions can be introduced from Equation~(\ref{E.15}) by
using the basic properties of the trace. For instance, as shown by
\citet{bol87}, there exists an infinite set of arbitrary
decompositions in terms of bound and scattering contributions of the
PL sum itself. So, as far as thermodynamic quantities are concerned,
only the total contribution of both bound and scattering states is
an unambiguous quantity. The $\rho^{5/2}$-contribution from bound
and scattering states merely reduces to its $\rho ^2$ counterpart
multiplied by $\beta e_\alpha e_\beta \kappa_D$. This
multiplicative factor arises from many-body effects which induce a
constant shift $-{e_\alpha }e_\beta \kappa_D$ on the energy levels
of the two-particles states.

Finally, the contribution $\rho ^2_\alpha E (-\sqrt{2}{{l_{\alpha
\alpha}} \over {\lambda_ {\alpha\alpha}}})$ arises from the exchange
of two charges $e_\alpha$ in the vacuum. It is finite, independently
of any screening effect, because the off-diagonal matrix elements $<
-\vec r {\,}\vert {\,}\exp \ ( - \beta h_{\alpha \alpha}) {\,}\vert
{\,}\vec r >$ are short-ranged. The magnitude of this contribution
is smaller than the one relative to free particles because the
repulsive potential $e^2_\alpha /r$ inhibits the exchange. Similarly
to what happens for the contributions of bound and scattering
states, at the order $\rho^{5/2}$, the many-body effects on the
two-particle exchange amounts to lowering the repulsive barrier
$e^2_{\alpha }/r$ by the constant $-e^2_\alpha {\,}\kappa_D$.

\subsection{The nature of the nonideal terms in the FK equation of state}

As is usual for a weakly nonideal plasma~\citep[see, {\it e.g.}][]{ebe76},
the most important contributions beyond the ideal gas limit are
the exchange terms, the Debye screening corrections, the short-range
corrections and diffraction (in decreasing order of importance). Let
us recall briefly the various conditions allowing us to check the
nature of the short-range corrections for a given temperature and
density.

Since the current level of our FK equation of state
is truncated at
the order $\rho^{5/2}$, it can only include processes with two-body effects.
It comes naturally, from the structure of the equation of
state, that short-range quantum effects are virtually decoupled
from the other corrections. This means that in all orders of density,
the main impact of these short-range quantum
corrections merely pertains to the
recombination and ionization processes.
The magnitude of the neglected higher-order terms will
indicate the influence of three-body or higher interactions.
The comparison with solar data will allow us
a quantitative estimate of these neglected terms.

Similarly, one has to be aware of the higher Debye corrections
beyond the $\rho^{5/2}$ term. A careful
analysis of the virial expansion shows that the quantum short-range
correction that only involves the two-body virial coefficient
$Q(x_{\alpha\beta})$ results in an alternating series in $\rho$, which
is therefore convergent. Since the leading Debye correction to pressure
(included in our formalism) is negative, the following term must be
positive. Should at some physical conditions the neglected higher corrections
lead to a lowering of pressure, we will be able to interpret them as a
manifestation of a recombination process occurring under these
thermodynamical conditions.

\section{Limits of validity of the FK equation of state}

\subsection{Dimensionless parameters}

A detailed analysis of this equation of state suggests rewriting
the truncated expression of the thermodynamic quantities (for
example pressure) in terms of three characteristic lengths:

$\bullet$ mean interparticle distance~~~~ $d_{\alpha}=(3/4\pi n_{\alpha})^{1/3}$

$\bullet$ thermal de Broglie wavelength~~~~~$\Lambda_{\alpha}=
(\beta h^2/m_{\alpha})^{1/2}$

$\bullet$ Landau length~~~~~~~~~~~~~~~~~~~~$l_{\alpha\beta}=\beta \ea\eb$

\noindent
Here the notation is standard, and there should be no confusion
between $\beta$ being both a label and $1/(kT)$. Note that since we are now
dealing with quantitative applications, we denote number density by
the usual symbol $n$ rather than the $\rho$ that was used in the
preceding theoretical sections of this paper.

Before the FK equation of state can be applied to the Sun,
we must determine where the formalism is applicable.
Here we evaluate the relevant criteria throughout the Sun, based on
a standard solar model
\citep[Model~S from Christensen--Dalsgaard, as described in][]{cd96}
First, we verify the well-known result that the plasma in the solar
interior is indeed not strongly coupled. Figure~\ref{fig1} shows the
coupling parameter

$$\Gamma_{\alpha} = {l_{\alpha\alpha} \over d_{\alpha}}$$

\noindent through the solar model. The index $\alpha$
denotes electrons, protons, and helium nuclei, respectively.
Nowhere does the coupling parameter $\Gamma_{\alpha}$ come near to
the critical value of unity.

Figures~\ref{fig2}-\ref{fig4} show the degeneracy parameter
$n_{\alpha} \Lambda_{\alpha}^3$, again for electrons, protons, and
helium nuclei, respectively.

\subsection{Qualitative discussion of ionization and recombination}

As we discussed in detail in section~2, avoiding too much
recombination is the most restrictive criterion for the FK equation of state.
Figures~\ref{fig5}-\ref{fig8} show that it is certainly more restrictive than coupling
and degeneracy. Obviously, our low-density virial expansion
cannot be used in the cooler outer layers of the Sun, where the plasma is far
from fully ionized. To obtain an estimate of limit of validity of FK,
which we identified with 90\% ionization, we have calculated the
hydrogen and helium ionization fraction using the
standard (ground-state-only) Saha equation

$${{N_{Z+1}N_{e^-}}\over{N_{Z}}} =
\bigl({{g_{Z+1}g_{e^-}}\over{g_{Z}}}\bigr)
{{V}\over{\Lambda_{\alpha}^3}}\exp(-{{\chi_Z}\over{kT}})\label{eqn:28}$$

\noindent
throughout the same solar model. As usual, $N_{\alpha}$ and $g_{\alpha}$
denote the number of particles and statistical weights of the various
species (here identified by the ion charge $Z$) and
$\chi_Z$ is the ionization potential pertaining to
the ionization reaction $Z \rightarrow Z+1$. Using this simple Saha equation for
this purpose is alright since we use it only for a kind of error analysis. And, as
numerous studies have shown, for solar conditions, this simple Saha equation is
everywhere at most 10\% wrong~\citep[~{\it e.g.},][]{cd92}.

Figures~\ref{fig5}-\ref{fig8} show the
recombination fraction for hydrogen, helium, and the first two
ionization stages of oxygen (as a typical representative heavy
element). The 90\%-ionization level is also shown.
Clearly, at some point in the colder outside layers, our
virial expansion must break down.

However, in the deeper layers of the Sun (and non-compact stars
in general), we can verify that the weak-coupling and low-degeneracy
criteria are fully satisfied and the limitation from the
``recombination criterion'' is totally negligible for the
temperatures and densities of those regions.

\subsection{The treatment of heavier elements}

Adding heavier elements (other than hydrogen and helium) reduces the domain
of validity of the FK equation of state to even higher temperatures. This is obvious
because it is harder to ionize these elements. Fig. \ref{fig8} shows this
quantitatively for oxygen.
We therefore adopt the simplest solution which is to ignore the heavier elements
altogether. Indeed, in the solar application
of Paper II, we will merely add the pressure of the heavier elements to the
FK expansion~\ref{E.14}. Since in the Sun, the number abundance of all
elements other than hydrogen and helium is only about $10^{-3}$, adding
the heavier elements to FK with an equation of state such as MHD, which
is at least as accurate as 1\% \citep[~{\it e.g.},][]{da06} will limit the total
error of such a simplified procedure to an acceptable order of magnitude of
$10^{-5}$.

In addition, knowing the domain of validity also allows us to study
selected recombination reactions in the framework of the FK equation
of state. For instance, in the calculation of nuclear reaction cross
sections, the ionization degree of the various nuclei plays a role,
because the statistical presence of electrons around the nucleus
modulates the reaction rate (enhancement factor). This has
consequences for solar physics especially with respect to the
neutrino problem, where the reactions involving {\rm Be} and {\rm B}
have to be know very accurately \citep{pc04}.

\subsection{Solar models with a hybrid equation of state}

Because of its limitations to nearly
full ionization, it is impossible to use the FK formalism for the entire Sun.
Therefore, for our applications in Paper II, we will have to use
a hybrid two-zone equation of state,
consisting of FK serving in the interior, and a more conventional one (OPAL or MHD)
in the outer layers. Although such a hybrid
equation of state will have discontinuities at the transition, these
discontinuities will not preclude the solution of the usual stellar structure
equations. Their main effect is expected to be a small jump in the gradient of the physical
quantities, similar to that caused, for instance, by
standard interpolation in equation-of-state
and opacity tables. As detailed studies~\citep{cd92} have shown, successful
equation of state tests can still be made despite such small imperfections.
The analysis of Paper II will reveal if the small discontinuities will
cause problems. If so, they can be avoided by the introduction of suitable
smooth ramp functions at the juncture line of the two-zone equation of state.

\subsection{Distribution of FK equation of state}

OPAL and MHD have been made available to solar and stellar modelers through
the use of tables. Although this saves the modelers a bit of computing
time, the tables have only been provided for a limited range of
conditions. In order to make the FK equation of state more versatile,
we plan to develop a computational routine for which users can define
their own input parameters such as element abundances. This
implementation of FK will be ideally suited for exploring issues such as He
diffusion and the inconsistency between seismology and the new values of heavy
element abundances \citep{asp05,ags05,gas07}.

\section{Conclusions}

Despite the limitation to nearly fully-ionized matter, the
FK approach can be applied to large parts of the solar interior,
more precisely to the part where helium is at least 90\% ionized.
In those regions the FK approach is applicable, and
it is especially suited to study
screening effects, bound states, recombination of fully ionized ions
and exchange effects.

Paper II will be dedicated to these observational tests.
Since the important relativistic correction for electrons~\citep{ek98}
is not part of the original FK equation of state~\ref{E.14}, we will add it in Paper II.
Paper III will go further,
on a rigorous theoretical level,
to the calculation of the thermodynamic functions
up to order $\rho^3$ in density. The additional higher-order terms will
allow us to take into account the three-body effects that occur
in a moderately ionized
plasma (for example, in the case of the solar center, the possibility of a
recombination of
${\rm He}^{++}$ is still an unsolved problem).

Despite the severe technical difficulties involving the practical application
of the FK equation of state for solar physics, there are unique features in the FK
approach that promise to turn it into the most exact of the available formalisms, if
they are limited to the layers of the Sun where more than 90\% of helium is
ionized. The localizing power of helioseismology allows us to test FK observationally,
promising both better and tighter constraints on the equation of
state. For instance, by virtue of its exactness,
the FK equation of state becomes a promising tool for tweaking parameterizable equations
of state such as MHD. More generally, the FK equation of state will lead to better
solar models as well as tighter solar constraints of the equation of
state.

\begin{acknowledgements}
We thank J\o rgen Christensen--Dalsgaard for the solar model used in
this study. This work was supported by the grant AST-0708568 of the
National Science Foundation.
\end{acknowledgements}

\end{document}